# Quantum Element Method for Simulation of Quantum Eigenvalue Problems

Ming-C. Cheng
Department of Electrical & Computer Engineering
Clarkson University
Potsdam, NY 13699-5720, USA

A previously developed quantum reduced-order model is revised and applied, together with the domain decomposition, to develop the quantum element method (QEM), a methodology for fast and accurate simulation of quantum eigenvalue problems. The concept of the QEM is to partition the simulation domain of a quantum eigenvalue problem into smaller subdomains that are referred to as *elements*. These elements could be the building blocks for quantum structures of interest. Each of the elements is projected onto a functional space represented by a reduced order model, which leads to a quantum Hamiltonian equation in the functional space for each element. The basis functions in this study is generated from proper orthogonal decomposition (POD). To construct a POD model for a large domain, these projected elements are combined together, and the interior penalty discontinuous Galerkin method is applied to stabilize the numerical solution and to achieve the interface continuity. The POD is able to optimize the basis functions (or POD modes) specifically tailored to the geometry and parametric variations of the problem and can therefore substantially reduce the degree of freedom (DoF) needed to solve the Schrödinger equation. The proposed multi-element POD model (or QEM) is demonstrated in several quantum-well structures with a focus on understanding how to achieve accurate prediction of WFs with a small numerical DoF. It has been shown that the QEM is able to achieve a substantial reduction in the DoF with a high accuracy beyond the conditions accounted for in the training of the POD modes.

## I. Introduction

A wide range of engineering and scientific analysis and design in the areas of electronics, photonics, materials, physics, biology, medicines and chemistry involves quantum eigenvalue problems governed by the Schrödinger equation [1-20]. These result from their small physical dimensions near the electron wavelength or molecular/nuclear scale, where quantum phenomena become essential and are primarily responsible for their photonic, electronic, and/or chemical characteristics. To understand complex behaviors in such small scales, the Schrödinger equation needs be solved, subjected to electric potential and energy band variations induced by, e.g., electric fields, photons, molecular structures, defects, etc. Simulation and design of semiconductor electronic and photonic nanodevices and nanostructures [1-7], which are the basic building blocks of modern computing and communication technologies, are one of the typical examples of the application of the Schrödinger equation. Efficiency and accuracy of simulation for multi-dimensional nanoscale electronic/photonic structures are crucial for continuing improvement in computing and communication technologies which we have been enjoying today at all levels of society.

Another example is the computations of electronic structures in nanomaterials and nanostructures based on density functional theory (DFT) for prediction of material properties and design of nanomaterials, drugs, biomaterials, etc. [10-20]. One of the important steps in the simulations using DFT is to solve the Kohn-Sham equations [21] for the electron wave functions (WFs). The WFs derived from the Kohn-Sham equations are basically determined by the $N$ single-particle Schrödinger-like equations subjected to a modified effective potential, representing interelectronic interactions, to account for many-body effects in the materials. The approximation that reduces a computationally unreachable many-body problem to an $N$ single-electron problem stems from the concept that the total energy of the system is a unique functional of the electron density. Even though the DFT makes it possible to perform multi-dimensional simulations of materials, it is still extremely computationally intensive, especially for large domain structures of materials with imperfections/defects [18-20].

To improve simulation efficiency of quantum structures, a projection-based reduced order model for the Schrödinger equation was developed previously [22] based on proper orthogonal decomposition (POD). The POD [23,24] is able to substantially minimize the number of numerical degrees of freedom (DoF) for simulation of a system of partial differential equations (PDEs) and has been successfully applied to some areas, such as fluid dynamics, heat conduction, microelectromechanical systems [23-36], etc. To develop a POD model, solution data for the PDE(s) of the domain structure subjected to parametric variations of interest are used to generate the optimal basis functions for the projected POD space. This is different from many other orthogonal basis methods using, for example, Fourier, wavelet, Legendre polynomials, Wannier functions, etc., in which the forms of the basis functions are selected based on the solution characteristics influenced by excitation sources and/or geometries. The POD is however similar to some projection-based machine learning algorithms, such as principal component analysis (PCA) [37-39] or singular value decomposition (SVD) [40-43] that also rely on data to generate and optimize the basis functions. The optimized (or *trained*) basis functions are therefore able to offer more accurate solution with a smaller DoF.

In the previous study of POD quantum eigenvalue problems [22], the electron WF in each quantum state (QS) is projected onto a POD space described by a finite number of basis functions (hereafter called POD modes). It has been demonstrated that the quantum POD model [22] offers very accurate solutions of the Schrödinger equation in multi-quantum well (multi-QW) structures using a very small number of numerical DoF (2 to 5 modes in general if the data quality is good), compared to direct numerical simulation. However, for multi-dimensional quantum eigenvalue problems, data collection and generation of the POD modes may be computationally intensive and become infeasible for a very large structure. One may also question the usefulness of the quantum POD model since, in order to construct a multi-dimensional quantum POD model, time-consuming numerical simulations need to be performed many times to collect WF solution data accounting for parametric variations induced by a range of excitations and/or boundary conditions.

To resolve this problem, a simulation methodology, termed the quantum element method (QEM), is proposed in this work, which implements the concept of reduced basis elements [44,45] in the quantum POD model [22]. The QEM combines domain decomposition with a reduced order model (the quantum POD model in this study), which partitions a large domain into smaller *subdomain blocks* (or *elements* hereafter). This approach offers several following advantages. First, computational resources to collect solution data for generating POD modes increase exponentially with the size of the physical domain. With smaller subdomains/elements, quantum

POD modes of each element can be generated more efficiently. Each element is projected onto a set of POD modes, and the projected elements can be *glued* together with the interface continuity enforced by the interior penalty discontinuous Galerkin method [51,52] to construct the simulation domain. Second, domain decomposition possesses a nature advantage for parallel computing and has been proven effective in parallel/distributed computing environments [46-50]. Finally, many quantum structures contain identical repeating substructures. Quantum POD modes of the generic elements can be generated and stored in a library of quantum elements, which can then offer cost-effective simulation/design of large quantum structures. The concept of the QEM is similar to a general practice in many engineering fields that rely on building blocks to deliver a cost-effective design process. Such a practice is popular not only for design of plumbing pipes and ventilation ducts in buildings but also for design of modern electronic and computing hardware, such as VLSI design or CPU/GPU architecture exploration utilizing standard cells or functional units.

The POD quantum approach developed previously [22] offers a single-element POD model for each QS. This poses a difficulty in the QEM when extending the QSs to the neighboring elements to construct a multi-element POD model (namely the QEM). A global POD model, which generates a set of POD modes for WFs in all the selected QSs, is thus developed first and applied, together with domain decomposition, to establish the methodology for the QEM. In this study, the developed methodology is applied to simple multi-QW structures with a focus on analysis of some detailed numerical aspects, including the least square (LS) errors and interface discontinuities of the predicted WFs associated with the number of modes and the penalty parameter, influences of the physical size and the complexity of the elements on the required number of modes to achieve a good accuracy, and robustness of the QEM beyond the training conditions for generating the POD modes. This first study of the QEM offers a fundamental understanding of the capability and limit of the proposed methodology. In the near future, this approach will be applied to multi-dimensional structures and to investigation of several other issues, such as further suppression of the interface discontinuity using methods other than the interior penalty discontinuous Galerkin method, influences of the penalty parameter on other quantum structures, error estimation, parallel computing, etc. POD for quantum value problems is presented in the next section with a demonstration on a single-element 6-QW structure. In Section III, a multi-element POD approach is presented to deliver the QEM. Demonstration of the developed QEM is presented in Section IV and final conclusions are drawn in Section V.

## II. Proper Orthogonal Decomposition for the Schrödinger Equation

POD generates a set of basis functions or POD modes from several sets of data samples that for the quantum eigenvalue problems may include wave functions (WFs) $\psi(\vec{r})$ of electrons or holes in nanomaterials or nanostructures. This is done by seeking a POD mode $\eta(\vec{r})$ that maximizes its mean square inner product using the data ensemble of the WFs,

$$\left\langle \left( \int_\Omega \psi(\vec{r}) \eta(\vec{r}) d\Omega \right)^2 \right\rangle \bigg/ \int_\Omega \eta(\vec{r})^2 d\Omega, \tag{1}$$

where $\Omega$ is the physical domain and the angled brackets $\langle\ \rangle$ indicate the average of the WF data ensemble collected over many sets of numerical observations accounting for the parametric variations. By finding a POD mode that maximizes (1), it implies that the component projected onto the POD mode $\eta(\vec{r})$ contains the maximum *L2* information of the system behavior described by the data ensemble. In the space orthogonal to this mode, the maximization process can be performed again to generate the second mode. Repetition of the maximization process in this fashion leads to an orthogonal set of POD modes.

Applying the variational calculus to the maximization process in (1), it can be shown [25,26] that this problem can be reformulated to the Fredholm equation of the second kind,

$$\int_{\Omega'} \mathbf{R}(\vec{r}, \vec{r}') \vec{\eta}(\vec{r}') d\vec{r}' = \lambda \vec{\eta}(\vec{r}), \tag{2}$$

where $\lambda$ is the POD eigenvalue of the data, representing the mean squared WFs captured by the corresponding POD mode, and $\mathbf{R}(\vec{r}, \vec{r}')$ is a two-point correlation tensor given as

$$\mathbf{R}(\vec{r}, \vec{r}') = \left\langle \vec{\psi}(\vec{r}) \otimes \vec{\psi}^T(\vec{r}') \right\rangle \tag{3}$$

with $\otimes$ as the tensor product. The decomposition process leads to an eigenvalue problem represented by (2) for a symmetric tensor $\mathbf{R}(\vec{r}, \vec{r}')$. Once the POD modes are found, the WF can be described by a linear combination of the POD modes,

$$\psi(\vec{r}) = \sum_{j=1}^{M} a_j \eta_j(\vec{r}), \tag{4}$$

where *M* is the selected number of modes representing the WF solution ($1 \leq M \leq N_s$ with $N_s$ as the number of sampled data sets), $a_j$ are weighting coefficients responding to the parametric variations, and the POD modes are normalized.

A set of equations for $a_j$ in (4) needed to determine the WF solution is obtained by projecting the Schrödinger equation onto the POD space described by the POD modes using the Galerkin projection method. The projected Schrödinger equation along the *i*th POD mode $\eta_i$ is given as

$$\int_{\Omega} \nabla \eta_i(\vec{r}) \cdot \frac{\hbar^2}{2m^*} \nabla \psi(\vec{r}) d\Omega - \int_{S} \eta_i(\vec{r}) \frac{\hbar^2}{2m^*} \nabla \psi(\vec{r}) \cdot \hat{n} dS + \int_{\Omega} \eta_i(\vec{r}) U(\vec{r}) \psi(\vec{r}) d\Omega = E \int_{\Omega} \eta_i(\vec{r}) \psi(\vec{r}) d\Omega, \tag{5}$$

where $\hbar$ is the reduced Planck's constant, $m^*$ is the electron/hole effective mass, *U* is the potential energy, *E* is the QS energy, and $\hat{n}$ is the outward normal vector of the boundary surface of the domain $\Omega$. The parametric variations are accounted for via *U*, which may be induced, for example, by external electric fields and/or the charge distributions in the structure. For simplicity, the WF near the boundary in the single-element domain is assumed small and the surface integral vanishes. In the multi-element structure, the surface integral in each element is coupled with the surrounding elements, which will be presented in Section III. Using (4) in (5), a matrix equation in POD space for $\vec{a}$ in terms of the *M*×*M* POD Hamiltonian matrix $\mathbf{H}_\eta$ can be derived ($M \leq Ns$),

$$\mathbf{H}_\eta \vec{a} = E_\eta \vec{a}, \tag{6}$$

where $\vec{a}$ becomes the eigenstate vector of $\mathbf{H}_\eta$ that is expressed as

$$\mathbf{H}_\eta = \mathbf{T}_\eta + \mathbf{U}_\eta \tag{7}$$

with the interior kinetic energy matrix given as

$$T_{\eta i,j} = \int_\Omega \nabla\eta_i(\vec{r}) \cdot \frac{\hbar^2}{2m^*} \nabla\eta_j(\vec{r}) d\Omega \tag{8}$$

and the potential energy matrix given as

$$U_{\eta i,j} = \int_\Omega \eta_i(\vec{r}) U(\vec{r}) \eta_j(\vec{r}) d\Omega. \tag{9}$$

Because the POD eigenvalue represents the mean squared WF captured by the corresponding POD mode, theoretically the least square (LS) error for an $M$-mode POD model can be estimated by

$$Err_{M,\lambda} = \sqrt{\sum_{i=M+1}^{N_s} \lambda_i \Big/ \sum_{i=1}^{N_s} \lambda_i}, \tag{10}$$

which predicts the number of modes needed to reach a desired accuracy if $N_s$ is large. Due to the limit of computer accuracy and also the inconsistency of numerical methods between the data collection and simulations, it may just provide an approximate error for the first several modes.

The decomposition procedure presented above can be applied to the quantum eigenvalue problems in two different ways. The POD can be performed separately for the data ensemble of WFs in each QS accounting for a range of parametric variations, which generates one set of POD eigenvalues and modes for each QS. In this case, (6) for each QS leads to $M$ POD eigenstates with only one physical state, which can be identified without difficulty [22]. An alternative is to perform the POD on WFs of all the selected $N_Q$ QSs to generate a global set of POD eigenvalues and modes for all selected $N_Q$ states. The later approach is investigated and demonstrated below in a single-element structure. The multi-element approach is presented in the next section. In all demonstrations in this paper, $Al_{0.3}Ga_{0.7}As/GaAs$ heterostructure is selected to construct QW structures. Electron effective mass $m^* = 0.0919 m_o$ in $Al_{0.3}Ga_{0.7}As$ and $m^* = 0.067 m_o$ in GaAs, and the conduction band discontinuity at the $Al_{0.3}Ga_{0.7}As/GaAs$ interface $\Delta E_C = 0.24$ eV.

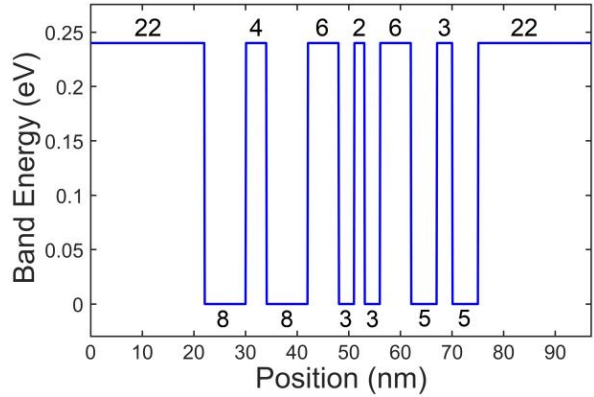

Fig. 1. An unbiased band diagram of the QW structure. The labeled numbers on the top indicate the barrier/spacer thicknesses and on the bottom the well widths.

In the global POD approach, $N_f$ distinct applied fields are applied to generated $N_s$ sets of WF data, where $N_s = N_f \times N_Q$ that offers a total of $N_s$ modes. $N_Q$-state WFs are collected at each applied field and $N_f$ should be greater than $N_Q$ to offer enough information for POD modes to successfully identify all $N_Q$ states if all the selected QSs are needed. The maximum allowed dimension of the POD Hamiltonian matrix in (6) is thus equal to $N_s$; however, a smaller dimension can be used. In the following demonstration $N_f$ is chosen as the dimension of the POD Hamiltonian, and a number of modes $M$ ($M \leq N_f$) is then selected to determine the WF in (4), depending on the desired accuracy. It should be pointed out that a dimension smaller than $N_f$ for the POD Hamiltonian can be used to solve (6), which will however reduce the dimension of $\vec{a}$ and thus the number modes available to predict the WF in each QS. It is however found that, if the number of applied fields $N_f$ is too much greater than $2N_Q$ (near $3N_Q$ for many QW structures we have examined), the global POD model actually produces unphysical/redundant QSs.

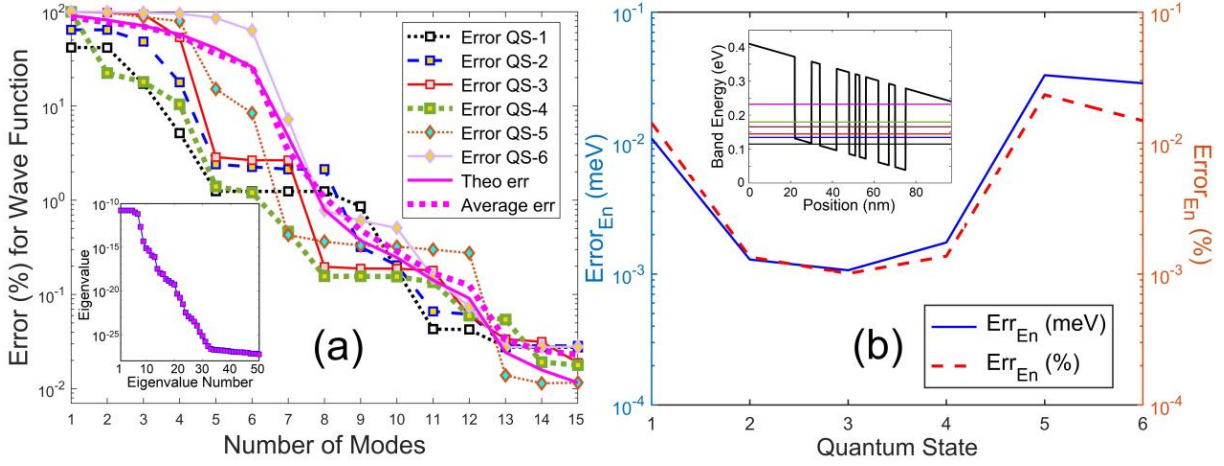

Fig. 2. (a) The LS error of the WF estimated from (12) in each QS as a function of the number of modes, together with the theoretical LS error $Err_{M,\lambda}$ estimated from (10). The eigenvalue Spectrum is shown in the inset. (b) The error of the predicted eigenenergy with an inset showing the eigenenergies of QSs 1-6 at -18 kV/cm. The percentage error is estimated with respect to the lowest band energy.

In the demonstration of the global POD model, 15 electric fields ($N_f$ = 15), ranging from -24 to 24 kV/cm with an equal division, are applied to a QW structure shown in Fig. 1 to collect 6 QSs ($N_Q$ = 6) for each field. This amounts to 90 sets of POD eigenvalues and modes ($N_s$ = 90). The eigenvalue spectrum for the first 50 modes is sketched in the inset of Fig. 2(a). The eigenvalue changes very little for the first 6 modes in order to account for essential information of the 6 QSs in the POD modes. It drops sharply after the 6th mode, and remains nearly unchanged beyond the 32nd mode, after decreasing over 16 orders of magnitude from the first mode, due to the computer precision. For the individual QS POD model presented in [22], the theoretical LS error $Err_{M,\lambda}$ in (10) is able to reasonably predict the numerical error $Err_M$ of the WF $\psi_{P,M}$ derived from the $M$-mode POD model estimated over the domain,

$$Err_M = \sqrt{\int_\Omega \left[\psi_{P,M}(\vec{r}) - \psi(\vec{r})\right]^2 d\Omega} \Big/ \sqrt{\int_\Omega \psi(\vec{r})^2 d\Omega}. \tag{11}$$

However, for the global approach because the eigenvalues are generated from WFs in all 6 QSs, $Err_{M,\lambda}$ in (10) is not able to predict the error in each state, as shown in Fig. 2(a). However, $Err_{M,\lambda}$ accurately predicts the average POD LS errors over all QSs that is defined as

$$Err_{M,ls,av} = \sqrt{\sum_{n=1}^{N_{QS}} Err_{M,ls,n}^2 \Big/ N_{NQ}} = \sqrt{\sum_{i=1}^{N_{QS}} \int_\Omega \left[\psi_{M,P,n} - \psi_n\right]^2 d\Omega \Big/ N_{QS}}, \tag{12}$$

where the $n$th state error for the $M$-mode POD WFs $Err_{M,ls,n}$ is defined in (11) with the integral of the WF square normalized to 1. As shown in Fig. 2(b), a maximum error near 0.023% (0.03 meV) is observed among all 6 state POD eigenenergies compared to the QS energies. The predicted 6 POD QS energy levels are also included in the inset of Fig. 2(b).

Based on the investigation of different QW structures, we have found that the first $N_Q$ modes always account for the essential information of the WFs in all the $N_Q$ states. Therefore, the eigenvalue of the first $N_Q$ modes remains nearly unchanged regardless of the selected number $N_Q$, as shown in the inset of Fig. 2(a) for the 6-QW case. Also, the first several modes in general contribute more to the lower QSs that thus need a smaller number of modes to reach an error near or below 1%, and the higher QS WFs (e.g., the 6th QS) needs 8 modes to offer an error near 1%. WFs in 4 QSs derived from the POD models with different number of modes are compared with the solution of the Schrödinger equation in Fig. 3.

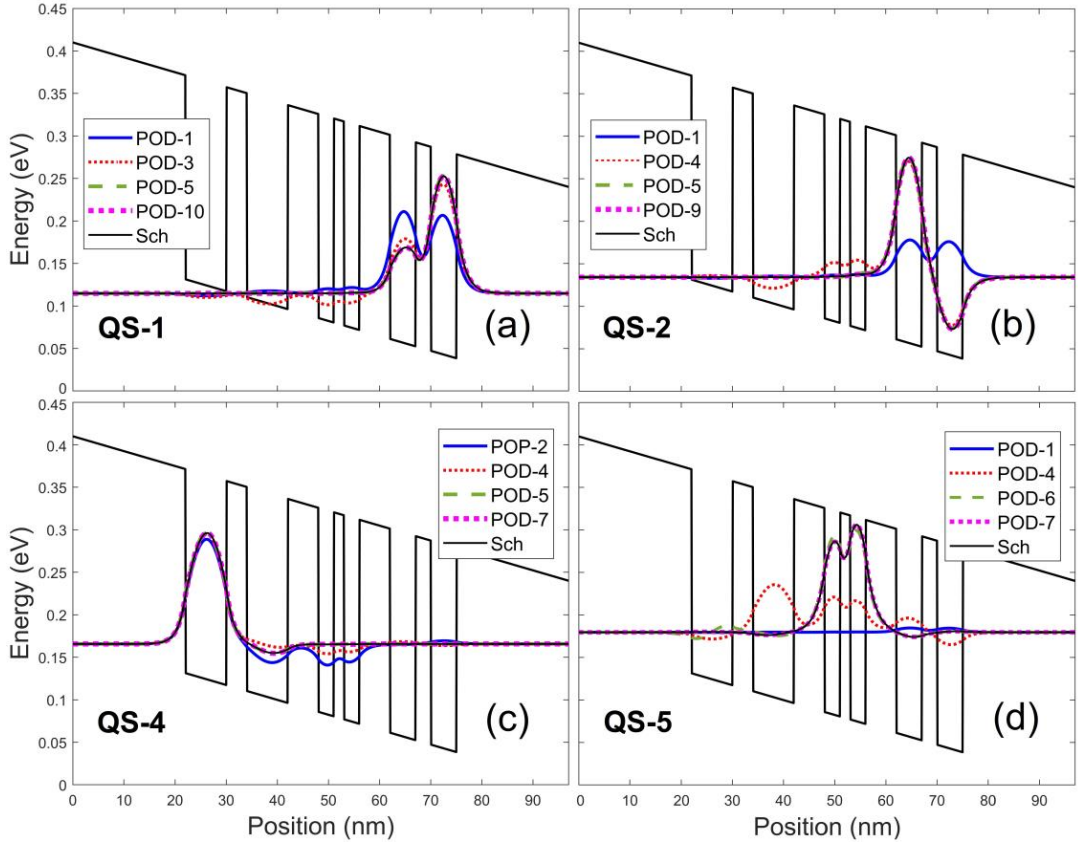

Fig. 3. Energy band diagram and WFs of four QSs in the 6-QW structure given in Fig. 1 at -18 kV/cm. Results are compared between the POD models and the Schrödinger equation.

### III. Quantum Element Method

For a very large quantum structure with fine meshes (thus a very large numerical DoF), it is impractical to apply the single-element approach presented in Section II or in [22] to generate POD modes for the whole structure. The concept of the domain decomposition is applied to partition the structure into multiple subdomains (or elements), and each selected element is projected onto its own POD space. These POD elements can then be glued together to construct a POD model representing the whole structure. The multi-element POD model or QEM thus projects the quantum eigenvalue problem of a large domain onto multiple sets of POD modes with each set representing an element in its POD space. Similar to the single-element approach, to arrive at a set of equations for the eigenvector $\vec{a}$ in a multi-element approach, (5) is modified to project the Schrödinger equation along the $i$th mode of the $p$th element,

$$\int_\Omega \nabla \eta_{p,i} \cdot \frac{\hbar^2}{2m_p^*} \nabla \psi_p d\Omega + \int_\Omega \eta_{p,i} U \psi_p d\Omega - \oint_S \eta_{p,i} \frac{\hbar^2}{2m_p^*} \nabla \psi_p \cdot \hat{n} dS = E \int_\Omega \eta_{p,i} \psi_p d\Omega. \tag{13}$$

Applying the interior penalty discontinuous Galerkin method [51,52] to the surfaces with all possible adjacent elements, (13) becomes

$$\int_{\Omega_p} \nabla \eta_{p,i} \cdot \frac{\hbar^2}{2m_p^*} \nabla \psi_p d\Omega - \sum_{q=1, q \neq p}^{N_{el}} \oint_{S_{pq}} \left[ \left[\!\left[ \frac{\hbar^2}{2m^*} \psi \right]\!\right]_{pq} \left\langle \nabla \eta_i \right\rangle_{pq} + \left\langle \frac{\hbar^2}{2m^*} \nabla \psi \right\rangle_{pq} \eta_{i\ pq} \right] \cdot d\vec{S}$$
$$+ \int_{\Omega_p} \eta_{p,i} U \psi_p d\Omega - \mu \sum_{q=1, q \neq p}^{N_{el}} \oint_{S_{pq}} \left[\!\left[ \frac{\hbar^2}{2m^*} \psi \right]\!\right]_{pq} \eta_{i\ pq} dS = E \int_{\Omega_p} \eta_{p,i} \psi_p d\Omega, \tag{14}$$

where $N_{el}$ is the total number of elements in the domain structure, $\mu$ is the penalty parameter defined as $N_\mu/dr$ with $dr$ as the local numerical mesh size and $N_\mu$ as the non-unit penalty number, and $[\![*]\!]_{p,q}$ and $\langle*\rangle_{p,q}$ are the difference and average, respectively, across the interface between the $p$th and $q$th elements. For any 2 elements that do not share an interface, the surface integrals in (14) vanish. A large positive value of $\mu$ is usually needed to stabilize the numerical result. There exists a minimum value $N_{\mu,min}$, that is independent of the local mesh sizes, to make the approach stable [51-55]. Different values of the penalty number $N_\mu$ have been used for different problems; for example, $N_\mu > 3$ was suggested in some problems [51-53] but some used 10 [55] or 20 [31,32,56]. This study however shows that $N_{\mu,min}$ is rather small for the quantum eigenvalue problem, which will be discussed in the following demonstrations.

(14) for the $p$th element projected along the $i$th mode can be rewritten as

$$\sum_{j=1}^{M_p} \left( T_{\eta_p,ij} + U_{\eta_p,ij} \right) a_{p,j} + \sum_{q=1, q \neq p}^{N_{el}} \sum_{j=1}^{M_p} B_{p,pq,ij} a_{p,j} + \sum_{q=1, q \neq p}^{N_{el}} \sum_{j=1}^{M_q} B_{pq,ij} a_{q,j} = E a_{p,i}, \tag{15}$$

where $M_p$ and $M_q$ are the selected numbers of modes in the $p$th and $q$th elements, respectively, the interior kinetic energy matrix for the $p$th element is given as

$$T_{\eta_p ij} = \int_{\Omega_p} \nabla \eta_{p,i} \cdot \frac{\hbar^2}{2m_p^*} \nabla \eta_{p,j} d\Omega, \tag{16}$$

the potential energy matrix as

$$U_{\eta_p ij} = \int_{\Omega_p} \eta_{p,i} U \eta_{p,j} d\Omega, \tag{17}$$

the diagonal boundary kinetic energy matrix as

$$B_{p,pq,ij} = -\frac{1}{2} \int_{S_{pq}} \frac{\hbar^2}{2m_p^*} \left[ (\nabla \eta_{p,i}) \eta_{p,j} + \eta_{p,i} (\nabla \eta_{p,j}) \right] \cdot d\vec{S} + \mu \int_{S_{pq}} \frac{\hbar^2}{2m_p^*} \eta_{p,i} \eta_{p,j} dS, \tag{18}$$

and the off-diagonal boundary kinetic energy matrix as

$$B_{pq,ij} = \frac{1}{2}\int_{S_{pq}} \frac{\hbar^2}{2m_q^*}\left[(\nabla\eta_{p,i})\eta_{q,j} - \eta_{p,i}(\nabla\eta_{q,j})\right]\cdot d\vec{S} - \mu\int_{S_{pq}} \frac{\hbar^2}{2m_q^*}\eta_{p,i}\eta_{q,j}dS. \tag{19}$$

From (15), a multi-element POD Hamiltonian matrix equation for the $N_{el}$-element domain can be expressed as

$$\begin{bmatrix} \mathbf{H}_1 & \mathbf{H}_{1,2} & \cdots & \mathbf{H}_{1,q} & \cdots & \mathbf{H}_{1,N_{el}-1} & \mathbf{H}_{1,N_{el}} \\ \mathbf{H}_{2,1} & \mathbf{H}_2 & \cdots & \mathbf{H}_{2,q} & \cdots & \mathbf{H}_{2,N_{el}-1} & \mathbf{H}_{2,N_{el}} \\ \vdots & \vdots & \ddots & \vdots & \iddots & \vdots & \vdots \\ \vdots & \vdots & \cdots & \mathbf{H}_p & \cdots & \vdots & \vdots \\ \vdots & \vdots & \iddots & \vdots & \ddots & \vdots & \vdots \\ \mathbf{H}_{N_{el}-1,1} & \mathbf{H}_{N_{el}-1,2} & \cdots & \mathbf{H}_{N_{el}-1,q} & \cdots & \mathbf{H}_{N_{el}-1} & \mathbf{H}_{N_{el}-1,N_{el}} \\ \mathbf{H}_{N_{el},1} & \mathbf{H}_{N_{el},2} & \cdots & \mathbf{H}_{N_{el},q} & \cdots & \mathbf{H}_{N_{el},N_{el}-1} & \mathbf{H}_{N_{el}} \end{bmatrix} \cdot \begin{bmatrix} \vec{a}_1 \\ \vec{a}_2 \\ \vdots \\ \vec{a}_q \\ \vdots \\ \vec{a}_{N_{el}-1} \\ \vec{a}_{N_{el}} \end{bmatrix} = E \begin{bmatrix} \vec{a}_1 \\ \vec{a}_2 \\ \vdots \\ \vec{a}_q \\ \vdots \\ \vec{a}_{N_{el}-1} \\ \vec{a}_{N_{el}} \end{bmatrix}, \tag{20}$$

where the entries of the diagonal block matrix $\mathbf{H}_p$ for the $p$th element are given as

$$H_{p,ij} = \sum_{j=1}^{M_p}\left(T_{\eta_p,ij} + U_{\eta_p,ij}\right) + \sum_{q=1,q\neq p}^{N_{el}}\sum_{j=1}^{M_p} B_{p,pq,ij} \tag{21}$$

and the entries of the off-diagonal block matrix $\mathbf{H}_{pq}$ is

$$H_{pq,ij} = \sum_{j=1}^{M_q} B_{pq,ij} \text{ with } p \neq q. \tag{22}$$

The WFs of the $p$th element are determined by its eigenvector $\vec{a}_p$ in (20) with a maximum number of modes equal to $M_p$. The WF in each state for the whole structure can then be constructed by combining WFs in space over $N_{el}$ elements based on (4) for each element. If $M_p$ modes is selected for the $p$th element, the matrix dimension of (20) is equal to

$$M_{N_{el}} = \sum_{p=1}^{N_{el}} M_p.$$

If the $p$th and $q$th elements do not neighbor each other, $B_{p,pq,ij} = 0$, $B_{pq,ij} = 0$ and thus $H_{pq,ij} = 0$. For a large number of POD elements, most of off-diagonal block matrices in (20) are zeros.

### IV. Demonstration of the Quantum Element Method

The QWs elements given in Fig. 4 are selected to construct POD elements for the QEM demonstration. 2 QW structures given in Figs. 5 (a) and 5(b) consisting of all the elements in Fig. 4 are used to collect WF data and generate POD modes of each element based on the approach presented in Section II. More specifically, simulations of the Schrödinger equation for the S-B-A-S structure in Fig. 5 (a) are used to collect the data of the first 6 QSs for Elements A, B, AS and SB, and simulations of the S-C-C-S structure in Fig. 5 (b) are for Elements C, SC and CS. 15 electric fields varying from -24 to +24 kV/cm are applied to the S-B-A-S structure and from -31 to +31 kV/cm to the S-C-C-S structure. The setups in Figs. 5 (a) and 5 (b) allow each element to experience different BCs induced by the adjacent elements as the field varies. In Fig. 5 (b), 2 possible sets of POD modes can be generated for Element C. The POD modes for Element C on the left are used in the following two demonstrations regardless of its neighboring elements. Also, Elements A and B in the demonstrations are neighbored by elements differently from those in Fig. 5 (a) during the training. As will be seen below, the POD modes appear to be quite robust. The modes for each element are able to capture the essential behavior of WFs even if the element is adjacent to other elements differently from those used in generation of its modes.

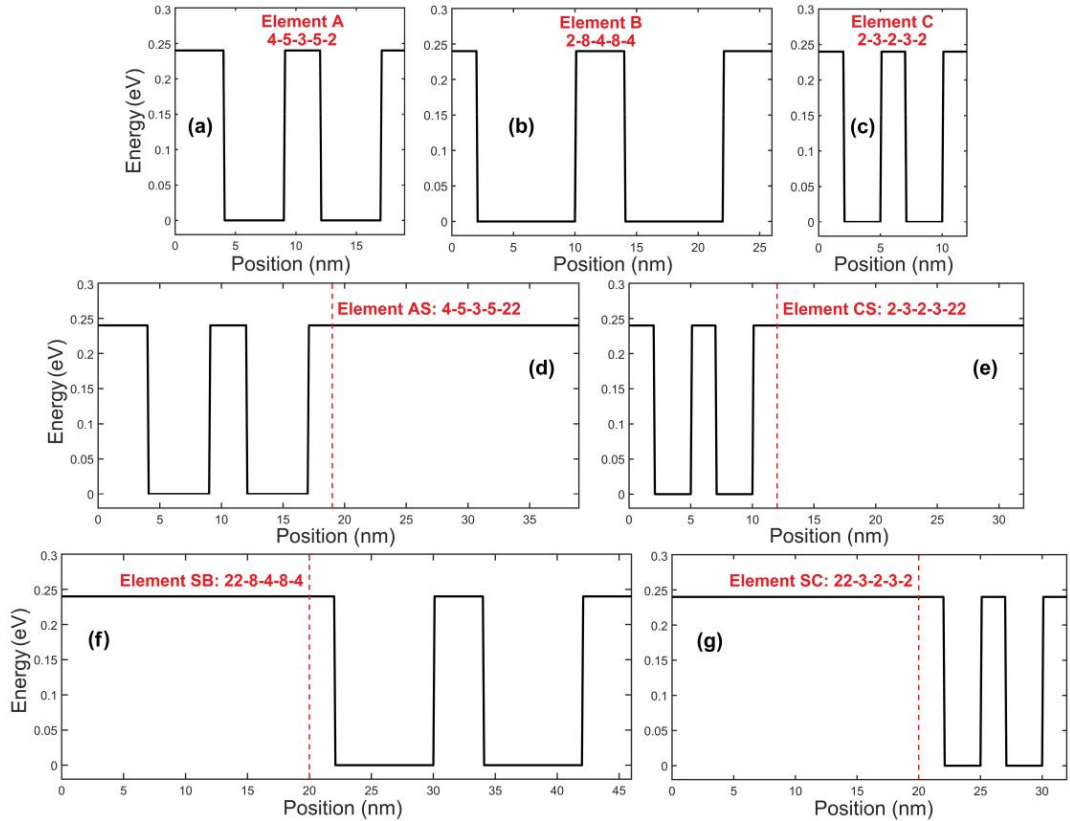

Fig. 4. QW elements used in the demonstrations of the QEM. The basic elements include Elements (a) A, (b) B and (c) C. A spacer with a thickness of 20nm is attached to the right sides of Elements A and C to construct Elements (d) AS and (e) CS, respectively. The spacer is attached to the left sides of Elements B and C to construct Elements (f) SB and (g) SC, respectively. The numbers included in each element indicate the thicknesses of the spacers/barriers and wells in sequence in nm.

The first demonstration of the QEM reexamines the 6-QW structure shown in Fig. 1 at the same electric field in the single-element simulation shown in Figs. 2 and 3. The 6-QW structure is constructed here by combining Elements SB, C and AS. With these projected POD elements, 3-element POD simulation with an equal number of modes in each element and $N_\mu = 20$ is performed. The LS errors and WFs are illustrated in Figs. 6 and 7, respectively. The 6 QS energies given in the inset of Figs. 6 predicted by this 3-element model is nearly identical to those shown in Fig. 2(b) from the single-element model. In general, a small error of WF (near or below 2%) can be reach with just 2 or 3 modes for each element except for higher states (QSs 5 and 6). Table I shows that the 3-element approach requires relatively less POD modes for each element than the single-element approach to reach a similar accuracy near or below 2% except QSs 5 and 6. This is because with a tilted band (non-zero applied electric field) in the multi-element structure the first several QS WFs are mainly confined in the lowest QSs in each element. As discovered in the single-element demonstration that the first several modes in general contribute more to the lower QSs whose POD models thus offer a good accuracy with a smaller number of modes. However, the total number of modes for the 3-element approach needed for a similar accuracy is still greater than that needed for the single-element approach. It should be noted that all these 3 elements neighbor elements differently from those used in generation of its POD modes.

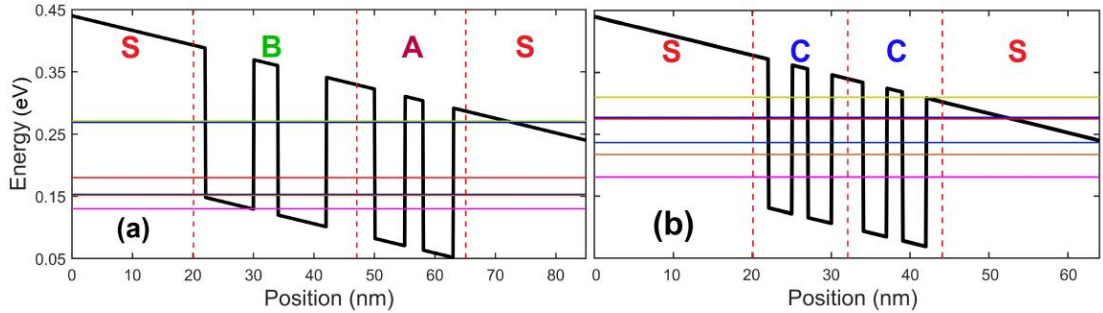

Fig. 5. Two structures selected to generate POD modes for each of Elements A, B, C, AS, CS, SB and SC. 15 electric fields ($N_s = 15$) with an equal division are applied, ranging from (a) -24 to +24 kV/cm and (b) -31 to +31 kV/cm. Also included are their energy band diagrams at its highest field and the 6 QS eigenenergies obstained from the Schrödinger equation. Each of the the following 2 eigenenergy levels are nearly degenerate: (a) $E_2$-$E_3$ and $E_5$-$E_6$ and (b) $E_5$-$E_6$.

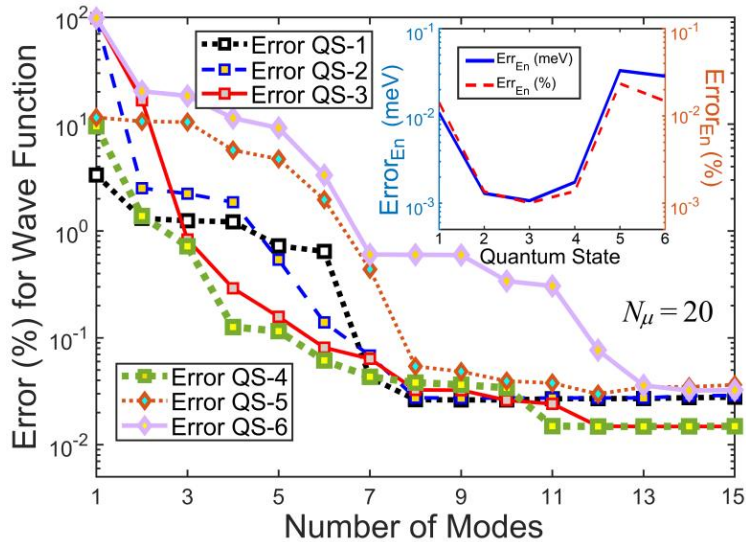

Fig. 6. LS errors of WFs of QSs 1-6 derived from the 3-element POD simulations of the 6-QW (SB-C-AS) structure at -18kV/cm.

Table I. Numbers of modes needed in each element to reach an error near or below 2%/1%/0.5% for the single- and three-element approaches in each QS

|        | QS-1   | QS-2  | QS-3  | QS-4  | QS-5  | QS-6   |
|--------|--------|-------|-------|-------|-------|--------|
| **Single** | 5/9/10 | 5/9/9 | 8/8/8 | 5/7/7 | 7/7/7 | 8/8/10 |
| **Three**  | 2/5/7  | 3/5/6 | 3/3/4 | 2/3/4 | 6/7/7 | 7/7/10 |

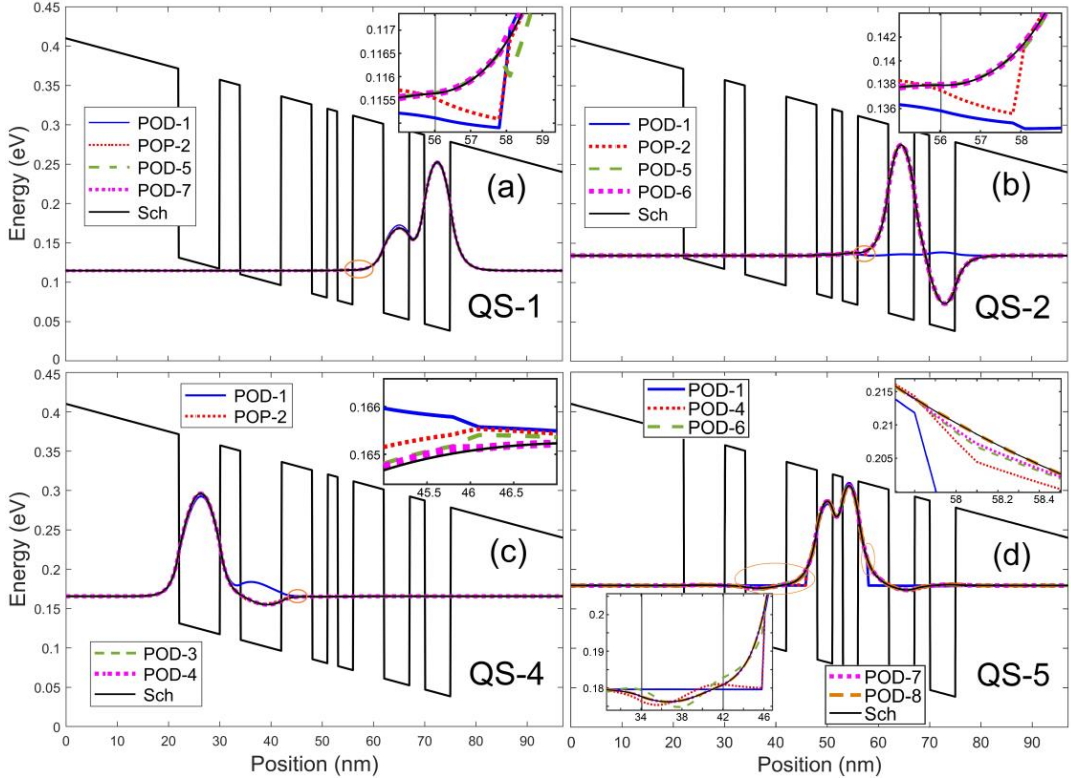

Fig. 7. Energy band diagram and WFs of four QSs in the 6-QW (SB-C-AS) structure at -18 kV/cm. Results are compared between the 3-element POD models and the Schrödinger equation.

WFs of QSs 1, 2, 4 and 5 illustrated in Fig. 7, compared to those shown in Fig. 3 derived from the single-element approach, reveal an interesting feature. In general, the 3-element POD model (Fig. 7) offers an accurate WF solution in each QS with a relatively smaller number of modes for each element than the single-element POD models (Figs. 2 and 3) except at interfaces between elements. For example, the QS-1 WF in Fig. 7(a) derived from the 3-element POD model and the Schrödinger equation are nearly on top of each other with just 2 modes (an LS error near 1%); it however needs 5 or 6 modes in Figs. 2(a) and 3(a). Similarly, it needs only 2 modes for each element to reach a high accuracy in QS 2, as shown in Fig. 7(b), but 4 and 5 modes are needed in Fig. 3. However, at the interfaces shown in the insets of Figs. 7(a) and 7(b), evident WF discontinuities are observed unless a larger number of modes are used. For example, 7 modes are needed to successfully suppress the discontinuity of the QS-1 WF shown in the inset of Fig. 7(a), which leads to a sudden drop in its error shown in Fig. 6 from 0.7% to 0.06% with the 7th mode added. In QS 2, the discontinuity shown in Fig. 7(b) becomes reasonably small with 5 modes and the error in Fig. 6 drops from 2% to 0.52% with the 5th mode included. With the 6th mode added, it brings the error down to 0.13%. Similarly, a significant reduction in the error due to the interfacial discontinuity is also observed in QS 4 when the 4th mode is included. For the 5th and 6th QSs, the WF discontinuities at both interfaces are relatively large and it requires more modes to minimize them. As shown in Figs. 6 and 7(d) for QS 5, when the discontinuity is successfully reduced from 6, 7 to 8 modes, the error drops from 2%, 0.44% to 0.055%, respectively.

In this 3-element structure, the LS errors in all 6 QSs remain nearly unchanged if $N_\mu > N_{\mu,min}$ with $N_{\mu,min} = 0.29$. When $N_\mu < N_{\mu,min}$, numerical instability occurs in some states, redundant states are generated and/or WFs in some states are identified incorrectly. For more complicated structures, selection of $N_\mu$ value actually has a strong influence on the accuracy of the predicted WFs in the higher QSs even if $N_\mu > N_{\mu,min}$. This will be demonstrated below.

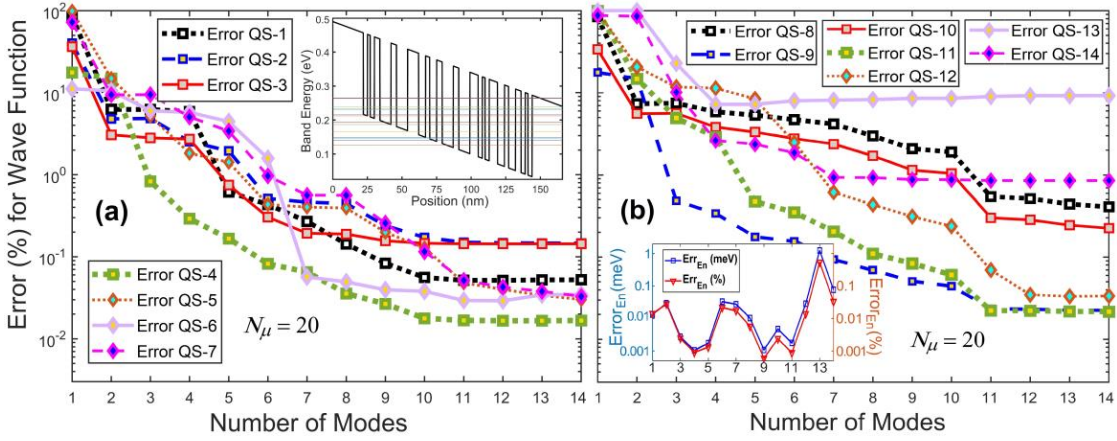

Fig. 8. LS errors of WFs in (a) QSs 1-7 and (b) QSs 8-14 derived from the 7-element POD simulations of the SC-B-A-B-C-A-CS structure at -15.06 kV/cm with $N_\mu = 20$. The insets show (a) the predicted 14 POD QS energies and (b) their errors.

To further verify the capability of the QEM and its accuracy influenced by the interface discontinuity, the QEM is applied to a 7-element QW structure consisting of Elements SC, B, A, B, C, A and CS in sequence using some of the POD elements in Figs. 4(a)-4(g). Simulations at applied voltages of 0.25V and 0.5V across this 166nm-long QW structure (electric fields of -15.06 and -30.12 kV/cm, respectively) are performed.

LS errors of the WFs in the 7-element structure derived from the QEM with $N_\mu = 20$ are displayed in Figs. 8(a) and 8(b) at -15.06kV/cm, together with the energy band diagram, predicted eigenenergies and errors of the POD eigenenergies included in the insets. The predicted WFs in some QSs are illustrated in Figs. 9(a)-9(f). Prediction of the QS energy is extremely accurate, as seen in the inset of Fig. 8(b), with an error less than 0.05% (0.1 meV) for all QSs except QS 13 whose error is near 0.6% (or slightly greater than 1 meV). A careful examination of Figs. 6-9 suggests that, if the predicted eigenenergy is very accurate in a QS (e.g., an error below 0.1%), the LS error of its WF drastically reduces once an enough number of modes are included to minimize the interface discontinuity. This is also observed in the next demonstration at higher electric field.

Fig. 9(a) shows that, when the discontinuity is minimized with 5 modes in QS 1, a sudden drop of its LS error is observed in Fig. 8(a). Except for QS 13, this is also observed in all other states where some need 3 to 6 modes to suppress the discontinuity (e.g., QSs 1-7, 9 and 11) but some need 11 modes (e.g., QSs 8 and 10). As shown in the insets of Fig. 9(d) for QS 8, large discontinuities at 77nm and 103nm lead to spurious ringing until they are minimized with the 11th mode added. A sudden drop in the error of the QS-8 WF is also observed in Fig. 8(b) when the 11th mode is included. In all the states but QSs 13 and 14, the LS error much less than 1% can be achieved $N_\mu = 20$ with an enough number of modes. Unlike other states, the QS-13 WF is the only WF that is not bounded and an evident error of the POD QS-13 WF with $N_\mu = 20$ is always observed even when the discontinuity is successfully suppressed with the 4th mode included; see Fig. 9(e). The LS errors of all the bounded WFs considerably smaller than 1% can be achieved except the QS-14 WF whose error 0.9%-1% with 7 or more modes in each element. It turns out, as illustrated in the inset of Fig. 9(f) near 130nm-165nm, there is a small probability for electrons to become unbounded beyond 150nm where the error percentage is very large regardless of the number of modes included. This explains why its LS error cannot be further reduced below 0.9%.

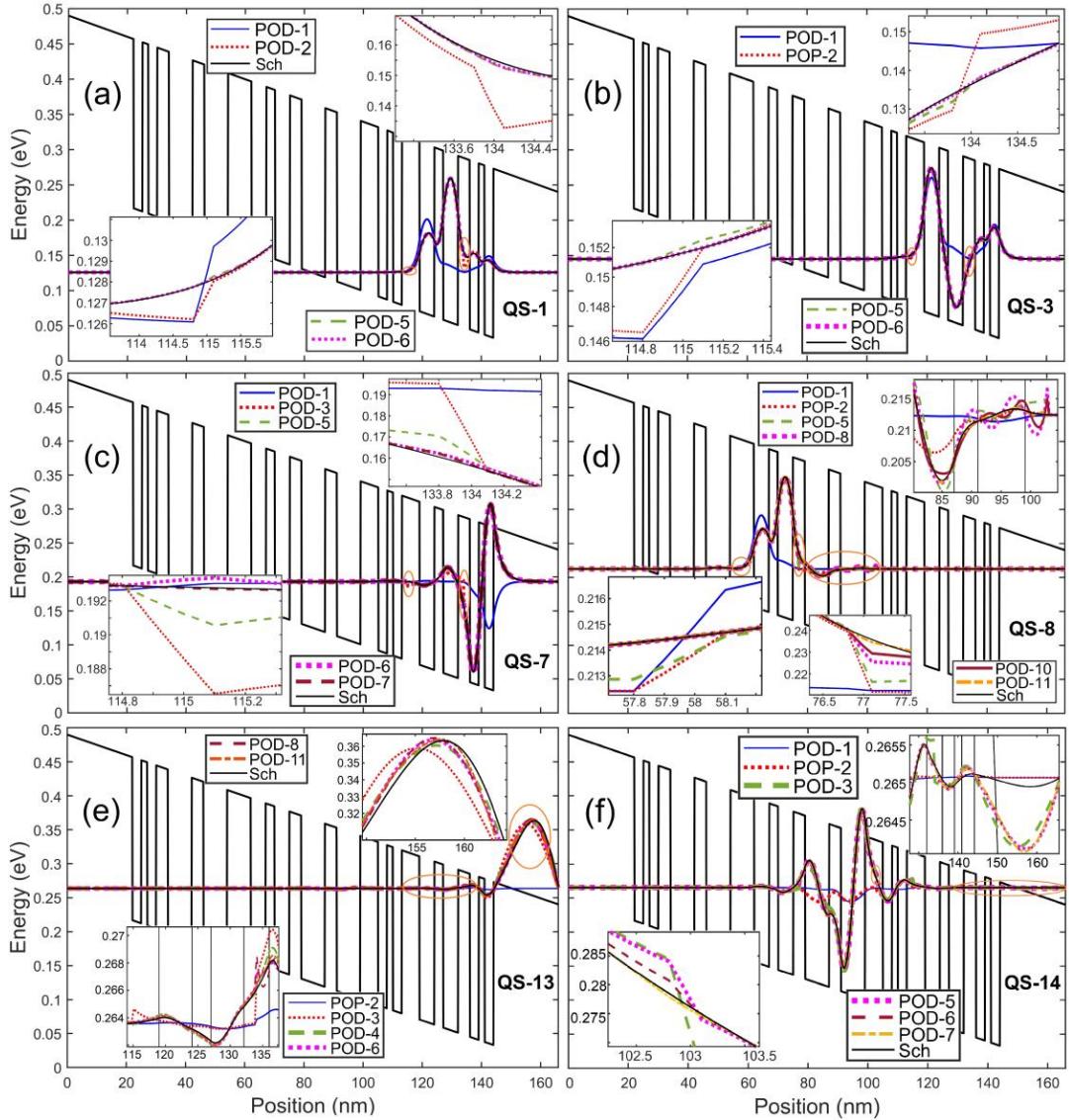

Fig. 9. Energy band diagram and WFs in the QW structure given in Fig. 15(a) at -15.06 kV/cm. Results are compared between the QEM with $N_\mu = 20$ and the Schrödinger equation.

Our study found that in this case with $N_\mu < 0.3$, numerical instability occurs in POD simulation and/or WFs cannot be identified correctly in some states. When $0.3 \leq N_\mu < 3.02$, a very small number of modes are able to offer very accurate WFs for the first 7 QSs; however, the POD model is not able to correctly predict the WFs in QSs 8-14 even with 15 modes for each element. When using $N_\mu \geq 0.302$, the LS errors of WFs in all states remain small and nearly unchanged except QSs 13 and 14 whose LS errors grow substantially with $N_\mu > 1$. The minimum LS errors for both QSs 13 and 14 become larger than 30% with 4 modes at $N_\mu = 32$. For $N_\mu > 32$, the POD model cannot correctly identify the WFs in QSs 13 and 14.

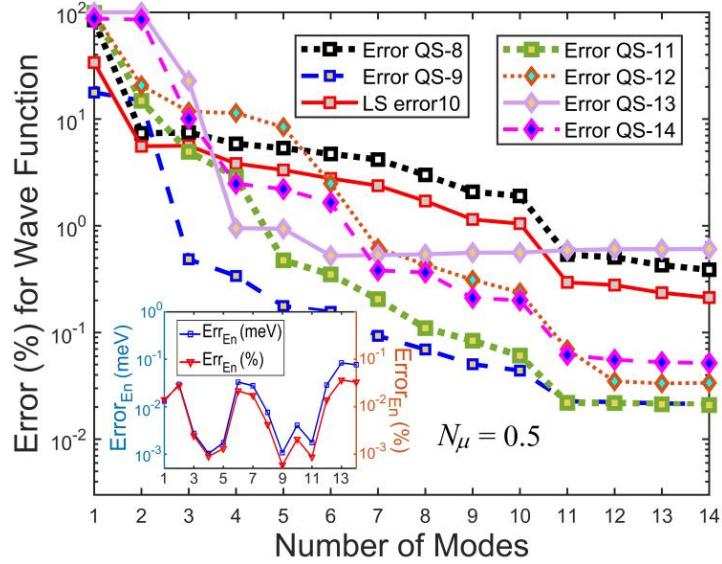

Fig. 10. LS errors of WFs in QSs 8-14 derived from the 7-element POD simulations of the SC-B-A-B-C-A-CS structure at -15.06 kV/cm with $N_\mu = 0.5$. The inset shows the predicted eigenenergies in QSs 1-14.

The LS errors of WFs in QSs 8-14 as functions of the number of modes with $N_\mu = 0.5$ are illustrated in Fig. 10, where errors of all the eigenenergies shown in the inset are considerably smaller than 0.1%. LS errors in Fig. 10 with $N_\mu = 0.5$ are nearly identical to those in Fig. 8(b) with $N_\mu = 20$ except for QSs 13 and 14. The LS errors in QS 13 with 4 modes and in QS 14 with 9 modes given in Fig. 10 drop nearly 80 and 50 times, respectively, and the errors drop more than one order with more modes, compared to those in Fig. 8(b). WFs in QSs 13 and 14 derived from the POD model with $N_\mu = 0.5$ are also displayed in Figs. 11(a) and 11(b) with insets showing the discontinuities at the same locations as those in Figs. 9(e) and 9(f). Use of $N_\mu = 0.5$ is able to successfully suppress the discontinuity in QS 13 with just 4 modes and it offers excellent prediction of the QS-13 WF. Although the discontinuities of the QS-14 near 103nm and 134 remain nearly unchanged when reducing $N_\mu$ from 20 to 0.5, significant improvement is observed in the predicted QS-14 WFs including its unbounded part. Results of this 7-element structure suggest that $N_{\mu,min} = 0.3$ for the first 7 QSs and 0.302 for the higher QSs. For a well-bounded WF, a high accuracy can be reached with a reasonably small number of modes if $N_\mu > N_{\mu,min}$. However, to achieve a better accuracy for the unbounded or not-well-bounded WFs, it is necessary to satisfy $N_{\mu,min} < N_\mu < 1$.

At 30.12 kV/cm, the large slope of the band diagram gives rise to more unbound states. Figs. 12(a) and 12(b) illustrate the LS error of the POD WF in each state as a function of the number of modes with $N_\mu = 10$. The predicted eigenenergies and their errors are also included in the insets. Based on the observation in Figs. 8(b) and 9(e), WFs in QSs 7, 9 and 11 are more likely to be unbounded because the errors in QSs 7, 9 and 11 are as large as 1%. This is indeed correct, as shown in Figs. 13(b) and 13(d) for WFs in QS-7 and QS-11 compared to their LS errors in Figs. 12(a) and 12(b). Even though the interfacial discontinuities of these 2 WFs can be successfully suppressed to some extent, their LS errors are still around 6%-16%. There is however one exception for the QS-13 WF whose error is as large as 8% but the error of its eigenenergy is only slightly above 0.01%. It turns out that the QS-13 WF is not well bounded in the QWs, as shown in Fig. 13(e). Except for these QSs whose WFs are not well bounded, LS errors of the predicted WFs in all other QSs are reduced significantly as soon as the interfacial discontinuities are improved, similar to Figs. 8 and 9 with $N_\mu = 20$. For example, when a much smaller discontinuity of the QS-3 WF is reached with 5 modes in Fig. 13(a), a sudden drop in its LS error is observed in Fig. 12(a). Similarly, a sudden error reduction of the QS-14 WF observed in Fig. 12(b) results from the improved discontinuity due to the added 7th mode shown in Fig. 20(f).

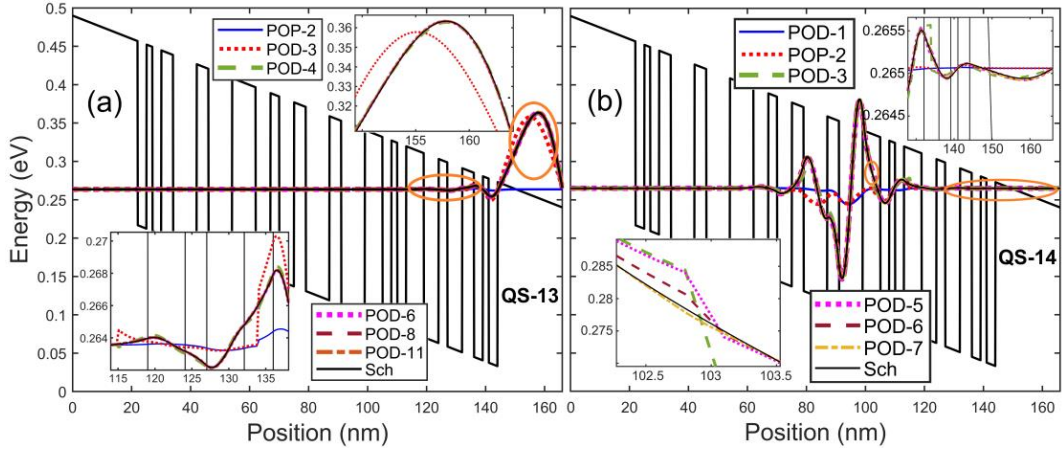

Fig. 11. Energy band diagram and WFs in QSs 13 and 14 of the QW structure given in Fig. 15(a) at -15.06 kV/cm. Results are compared between the QEM with $N_\mu = 0.5$ and the Schrödinger equation.

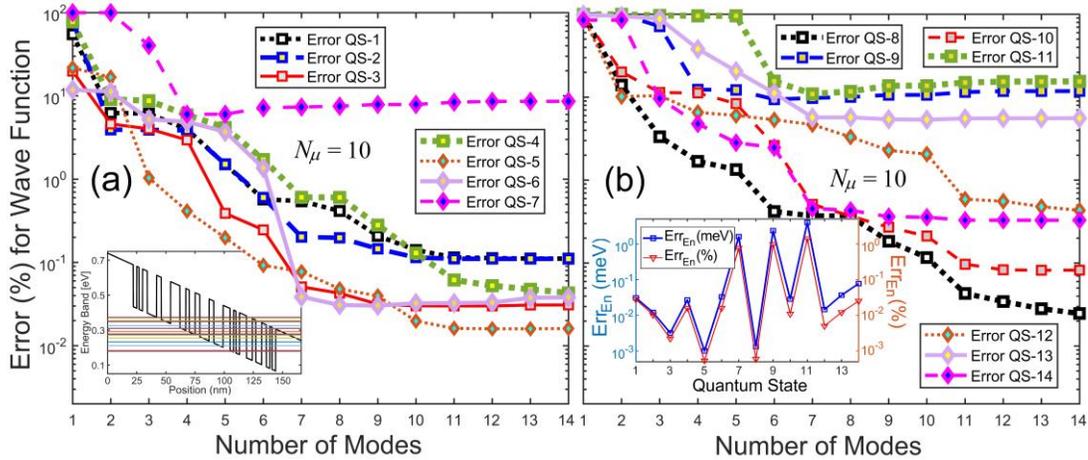

Fig. 12. LS errors of WFs in (a) QSs 1-7 and (b) QSs 8-14 derived from the 7-element POD simulations of the SC-B-A-B-C-A-CS structure at -30.12 kV/cm with $N_\mu = 10$. The insets show (a) the predicted 14 QS energies in the energy band diagram and (b) their errors.

For this structure at 30.12 kV/cm, the value of $N_{\mu,min}$ is actually the same as that at 15.06 kV/cm; namely $N_{\mu,min} = 0.3$ for the first 7 QSs but $N_{\mu,min} = 0.302$ for QSs 8-14. A very small number of modes are able to offer a good accuracy for QSs 1-6 for $N_\mu > 0.3$ regardless how large $N_\mu$ is. However, as $N_\mu > 1$, the LS errors of WFs in QSs 7-14 increase evidently when the same number of modes are used, especially for the QS-12 WF and all the unbounded WFs (QSs 7, 9 11 and 13). As $N_\mu > 22$, the POD model is not able to correctly identify WFs in QSs 11 and 12 and the smallest LS errors in QSs 7, 9 and 13 stay nearly unchanged around 10%-20% with 7 or more modes.

The LS error of the WFs predicted by the POD model with $N_\mu = 0.5$ as a function of the number of modes is illustrated in Fig. 14 that only includes the QSs whose LS errors reduce evidently compared to Figs. 12(a) and 12(b) with $N_\mu = 10$. The inset also displays the error of the eigenenergy in each state, where all errors are below 0.2%. As $N_\mu$ decreases from 10 to 0.5, LS errors of the bounded WFs (QSs 10 and 14) are only decreased by 2 - 3 times with 11 or more modes; the errors however drop by nearly an order of magnitude for the unbounded WFs (QSs 7, 9, 11 and 13). The POD WFs in QSs 7 and 11 with $N_\mu = 0.5$ are also shown in Figs. 15(a) and 15(b) with insets showing the discontinuities at the same locations as those in Figs. 13(b) and 13(d) using $N_\mu = 10$. When using $N_\mu = 0.5$, significantly improvements are observed for the discontinuity in QS 7 near 134nm with 5 or more modes and for the unbounded WF beyond 150nm with 4 or more modes. In QS 11, although the discontinuity near 134nm is not improved much until the 10 mode is included, the errors of the WF over the whole structure is substantially reduced, especially near 125-135nm, as shown in the inset of Fig. 15(b) compared to Fig. 13(d).

The findings from the above 3 demonstrations of the QEM are summarized as follows. To stabilize the multi-element POD model, $N_{\mu,min}$ in (14) is approximately 0.3 for the lower level QSs and slightly higher for the higher QSs. For lower QSs (QSs 1-6 in our cases of study), as long as $N_\mu > N_{\mu,min}$, a small number of modes are able to offer a high accuracy in the predicted WF and its LS error is nearly independent of $N_\mu$. To achieve a good accuracy for the unbounded WF, $N_\mu$ needs to stay within the range of $N_{\mu,min} < N_\mu < 1$. For the bounded WFs in higher QSs, most likely the POD results are accurate if $N_\mu > N_{\mu,min}$; however some exceptions were observed, such as QS 12 at 30.12 kV/cm for $N_\mu > 22$ and QS14 at 15.06kV/cm for $N_\mu > 32$ in the 7-element structure. To ensure the accuracy of the QEM, it would be a good practice to select $N_{\mu,min} < N_\mu < 1$. It is also observed that use of $N_{\mu,min} < N_\mu < 1$ tends to minimize some interface discontinuities of the unbounded WFs and give rise to a relatively smaller LS error. In the worst cases among all well-bounded higher-QS WFs, an LS error below 0.6% can be reached if 11 modes in each element are used in the 7-element structure at both voltages, as shown in Figs. 8(b) and 12(b), where $N_\mu = 20$ and 10, respectively. Even in these worst cases, reduction of the DoF offered by the QEM is still more than one order of magnitude, compared to numerical simulation of the Schrödinger equation, if one desires to accurately predict nearly degenerate eigenenergy levels. Also, only 7 modes for each element are needed to achieve an accuracy of 0.5% for the lower-QS WFs (QSs 1-6). In this study an equal number of modes are used in all elements. Further reduction in DoF is expected if an unequal number of modes are used in each element.

The applications of the QEM has clearly demonstrated that the QEM is more robust than expected. The trained POD modes are able to accurately predict the well-bounded WFs with $N_\mu > N_{\mu,min}$ even when most elements in the structure are neighbored by those different from the elements used in the training process. For the unbounded WFs, a good accuracy can be reached if $N_{\mu,min} < N_\mu < 1$. In addition, Elements A and B (see Fig. 12) only experienced electric fields as high as 24 kV/cm during the training but in one of demonstrations a field higher than 30 kV/cm is applied. Regardless of the conditions significantly different from how the POD modes are generated, these modes are somehow capable of predicting the WFs accurately with a DoF considerably smaller than that needed in numerical simulation of the Schrödinger equation.

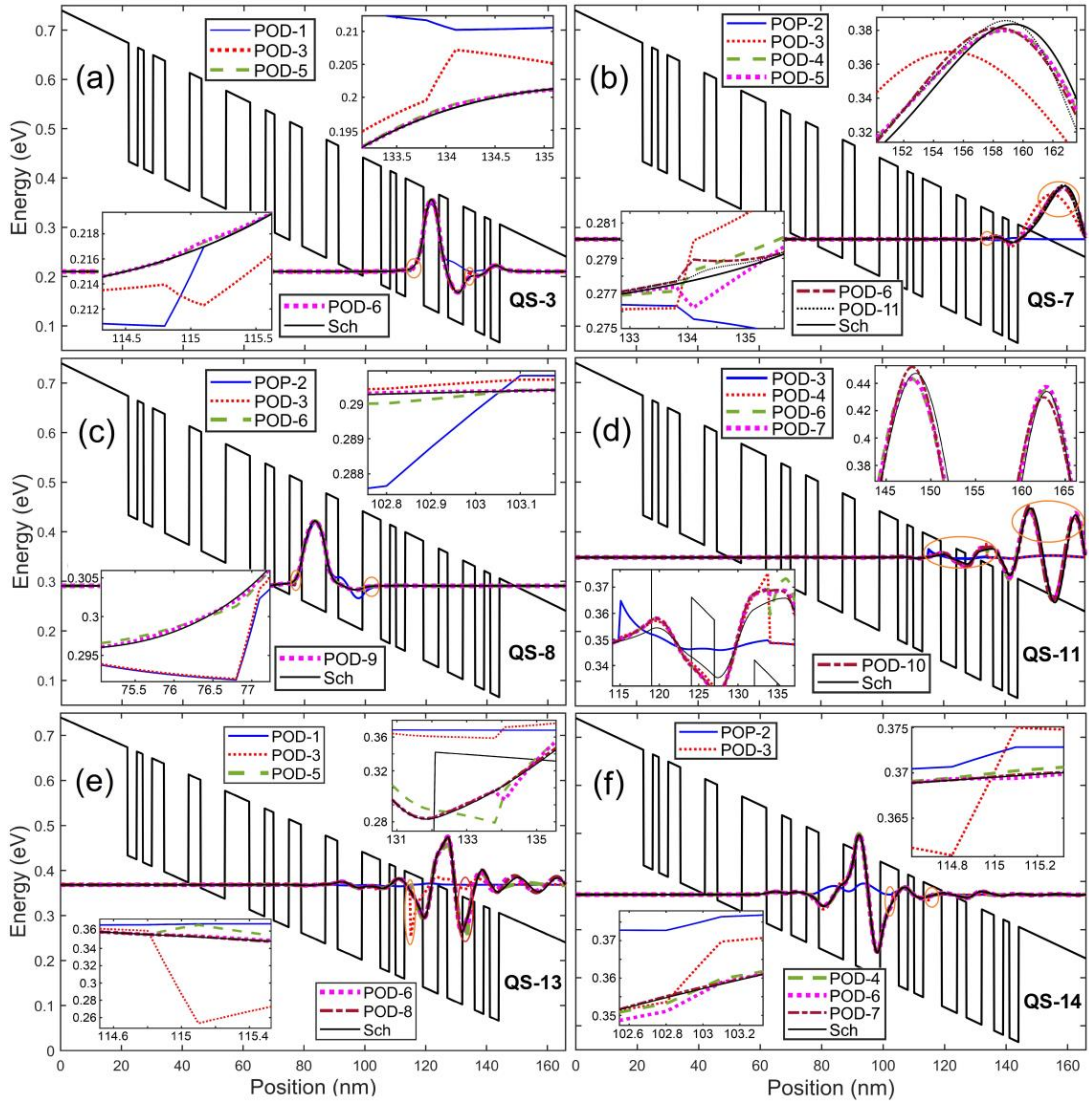

Fig. 13. Energy band diagram and WFs in the QW structure given Fig. 12(a) at -30.12 kV/cm. Results are compared between the QEM with $N_\mu = 10$ and the Schrödinger equation.

An interesting feature is also observed in the demonstrations. The number of modes needed for each element to reach a similar accuracy in QSs do not vary much with the element size or the number of QWs. At least this is the case in this study for first 6 QS WFs. For example, in order to reach an error near 1% in Fig. 2, the single-element structure with 6 QWs needs approximately 8 modes. In the 3-element and 7-element structures presented in Figs. 6, 8 and 12 with 2 QWs in each element, it requires 6 or 7 modes for each element to reach a similar accuracy. The number of QWs in the one element structure in Fig. 2 is 3 times as large as that in each element in Figs. 6, 8 and 12. Also, the length of the total QWs in one element (excluding the spacers) in Fig. 2 is 2 to 5 times as large as that in each element of Figs, 6, 8 and 12. Despite the substantially more complicated and larger structures, the number of modes for each element needed in Fig. 2 is only 15%-33% larger. It is thus more desirable to select larger-size elements to construct a large domain structure in order to reduce the DoF in the POD simulation. However, the larger the element size is, the more intensive computation and larger memory space are needed for collecting the WF data to generate the POD modes. The trade-off between these 2 factors needs to be considered.

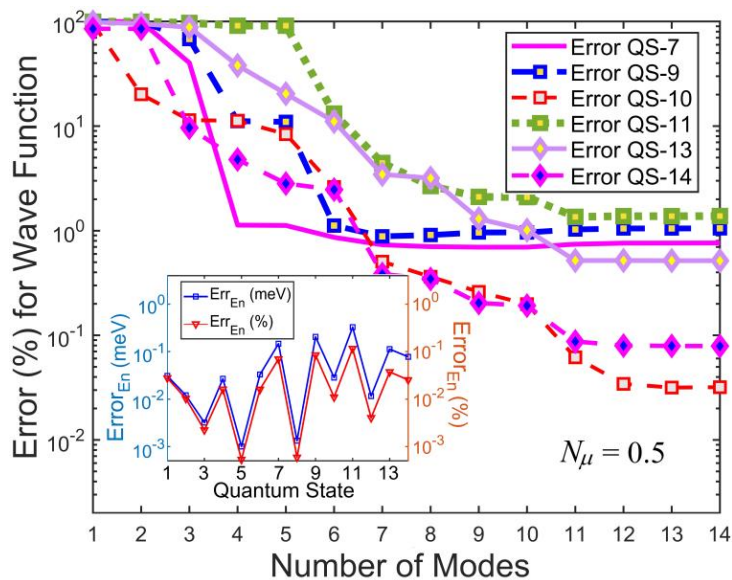

Fig. 14. LS errors of WFs in some higher QSs from the 7-element POD simulations of the SC-B-A-B-C-A-CS structure at -30.12 kV/cm with $N_\mu = 0.5$. The inset shows the predicted eigenenergies in QSs 1-14.

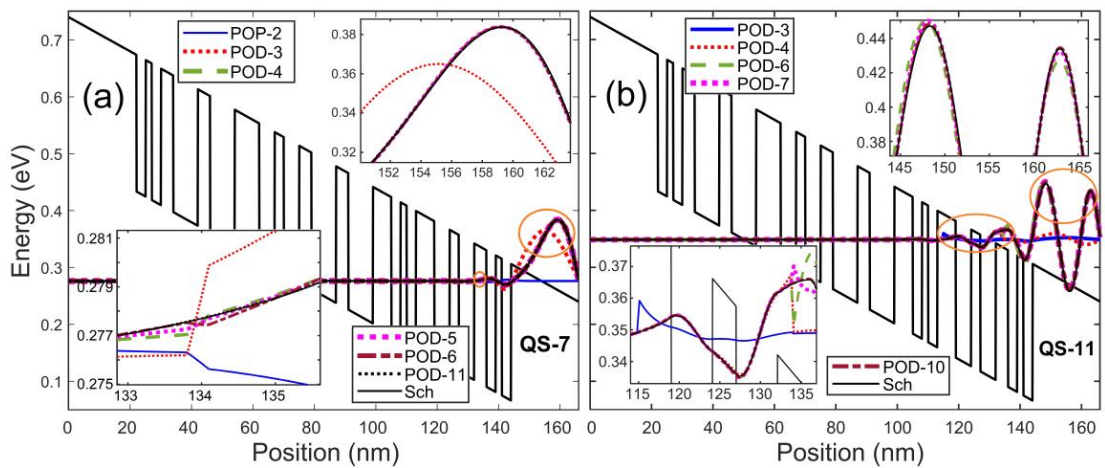

Fig. 15. Energy band diagram and WFs in QSs 7 and 11 of the QW structure given in Fig. 12(a) at -30.12 kV/cm. Results are compared between the QEM with $N_\mu = 0.5$ and the Schrödinger equation.

## V. Conclusions

A quantum element method (QEM) has been presented, which combines the quantum POD model [22] with the concept of domain decomposition to deliver a multi-element POD methodology. In order to develop the POD methodology for the quantum eigenvalue problems, the single-element global POD model was presented first and then applied to develop the QEM. Demonstrations of the QEM show that accurate WFs with a very small DoF can be achieved for the first several QSs in QW structures if the penalty number $N_\mu > N_{\mu,min}$. However, for higher QS WFs and unbounded WFs a larger DoF is needed to effectively minimize the interface discontinuities between elements to reach a good accuracy if $N_{\mu,min} < N_\mu < 1$.

This study reveals some encouraging capabilities in the proposed QEM. The generated POD modes, once coupled with the Schrödinger equation, are able to predict the well-bounded WFs with a high accuracy and the unbounded WFs with a good accuracy (with an error near 1%) in the conditions that were not considered during the data collection/training process, such as higher electric fields and different neighboring elements. This capability allows more flexible selection of the generic elements that can be stored in a library of quantum elements for design/simulation of large quantum structures. Results observed in this study also suggest that the number of modes needed for each element to reach a similar accuracy in QSs is not sensitive to the size of the element or the complexity of the structure. The larger elements are therefore desirable to further reduce the DoF in the multi-element POD simulation of a large-domain quantum structure. Overall, in the worst case scenario for the well-bounded WFs illustrated in the demonstrations with the conditions beyond how the modes are generated/trained, the QEM is able to achieve a reduction in the DoF over an order of magnitude with an LS error below 0.6%, compared to numerical simulation of the Schrödinger equation.

This work presented the first investigation of the QEM based on the proposed multi-element POD methodology for quantum eigenvalue problems. Demonstrations of the proposed methodology in some QW structures have offered information on the capability and limitation of the approach. There are still many issues that need to be investigated, including alternative schemes to more effectively minimize the interface discontinuity, parametric variations in addition to the changes in the potential slope, parallel computing for the QEM, etc. This study however offers an innovative concept that may eventually lead to a powerful approach to efficient large-scale simulations of quantum eigenvalue problems. The proposed approach could be particularly useful for simulation of quantum structures that require intensive computation, such as nanostructures or materials with localized defects or imperfections.